\pdfoutput=1
\documentclass[conference,a4paper]{IEEEtran}
\usepackage{graphicx}
\begin{document}
\title{Using solar and load predictions in battery scheduling at the residential level}
\author{\IEEEauthorblockN{Richard Bean}
\IEEEauthorblockA{Redback Technologies\\
	Brisbane, Australia}
\and
\IEEEauthorblockN{Hina Khan}
\IEEEauthorblockA{School of ITEE\\
The University of Queensland, Australia}
}
\maketitle
\begin{abstract}
Smart solar inverters can be used to store, monitor and manage a home\textquoteright s solar energy. We describe a smart solar inverter system with battery which can either operate in an automatic mode or receive commands over a network to charge and discharge at a given rate. In order to make battery storage financially viable and advantageous to the consumers, effective battery scheduling algorithms can be employed. Particularly, when time-of-use tariffs are in effect in the region of the inverter, it is possible in some cases to schedule the battery to save money for the individual customer, compared to the ``automatic'' mode. Hence, this paper presents and evaluates the performance of a novel battery scheduling algorithm for residential consumers of solar energy. The proposed battery scheduling algorithm optimizes the cost of electricity over next $24$ hours for residential consumers. The cost minimization is realized by controlling the charging/discharging of battery storage system based on the predictions for load and solar power generation values. The scheduling problem is formulated as a linear programming problem. We performed computer simulations over 83 inverters using several months of hourly load and PV data. The simulation results indicate that key factors affecting the viability of optimization are the tariffs and the PV to Load ratio at each inverter. Depending on the tariff, savings of between 1\% and 10\% can be expected over the automatic approach. The prediction approach used in this paper is also shown to out-perform basic “persistence” forecasting approaches. We have also examined the approaches for improving the prediction accuracy and optimization effectiveness.
\end{abstract}
\IEEEpeerreviewmaketitle
\section{Introduction}
\begin{figure}[t]
	\centering
	\includegraphics[width=6.0cm]{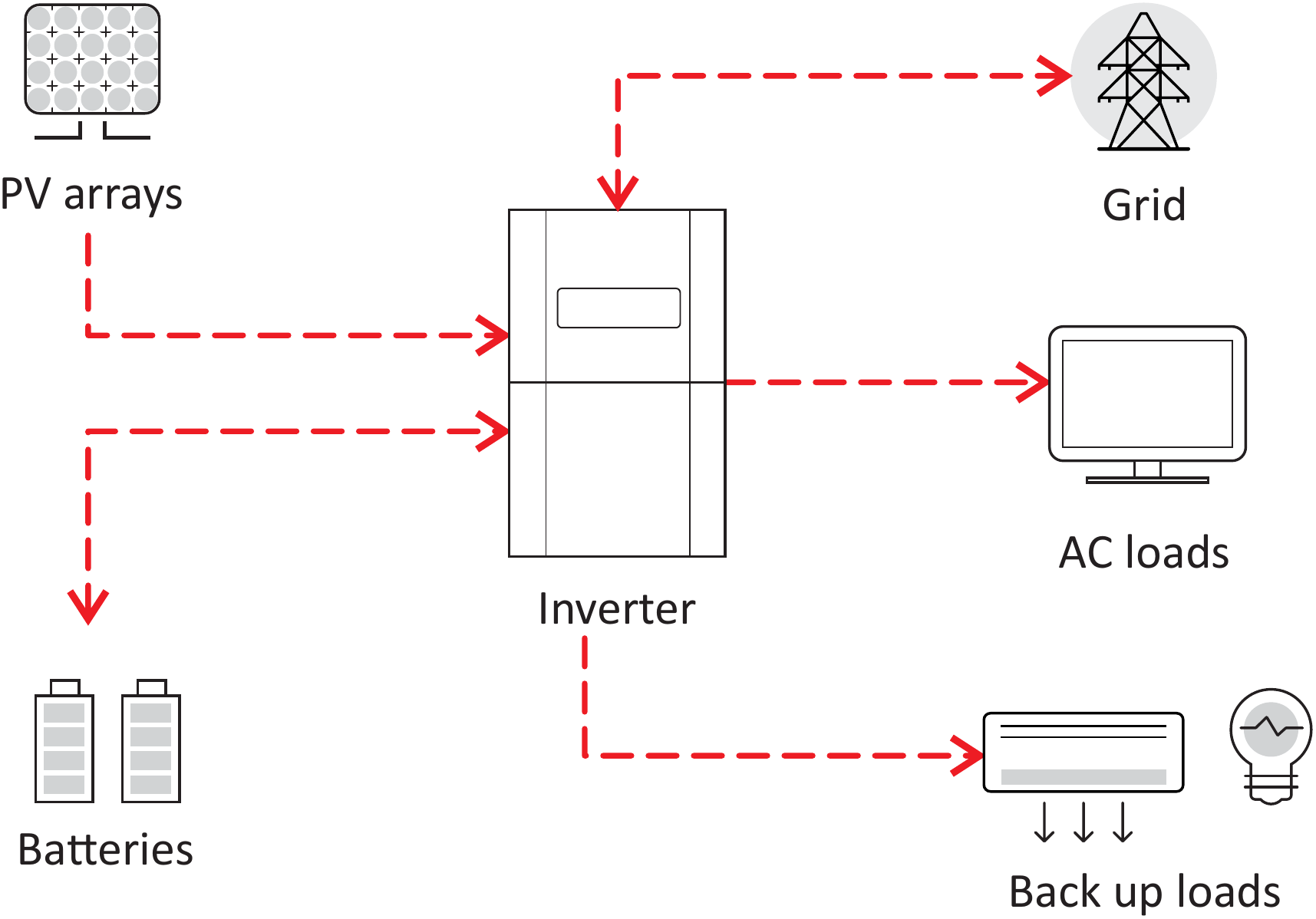}
	\caption{Schematic of inverter with associated electrical loads, battery and grid connections}
	\label{fig:snap}	
	\vspace{-11pt}
\end{figure}

Solar inverters are devices which transform solar radiation into AC or DC power. This power is then used immediately to meet electrical load, stored in a battery or storage device, or sold to a grid operator. Power can also flow through the inverter from the battery or grid to meet load. These flows of power can be seen in Figure \ref{fig:snap}. This technology will form an essential part of reducing carbon emissions worldwide to meet Paris Climate Agreement goals and can also provide energy security at the household level while increasing the load factor (that is, the ratio of average load to peak load) at the grid level.
Redback Technologies is an Australian manufacturer of {\it smart} solar inverters. Unlike traditional inverters, the {\it smart} inverters are able to send and receive messages quickly, as well as share granular data with the owner, utility and other stakeholders. Such inverters can store, monitor and manage a home\textquoteright s solar energy. Within a home, appliances can be attached as ``AC loads'' or ``Backup loads''. Backup loads can be run from the inverter battery and solar power when grid power is not available. Typically, the total load for the home has an instantaneous value less than 10 kW. Most homes have an average load of at least 300 W. The instantaneous photovoltaic power is denoted by PV; the solar panels attached to the inverter are of the order of 5 kW. If the inverter is located in Australia with 5 kW of solar panels, this corresponds to an average daily output of between 17.5 kWh (Hobart) and 25 kWh (Alice Springs) (\cite{solarchoice}).
The batteries attached to the inverter have an associated state of charge value which must be kept between a range of values (e.g. 20\%-100\%) to avoid adverse effects or battery failure. The information related to the load, PV, state of charge and health of the battery is stored in a central location and is updated regularly. This enables users to monitor and manage their energy needs from a smart phone application. The users can view the cumulative historical values as well as instantaneous values. The predictions for PV and Load can also be viewed. If tariffs are known the user can also see estimated costs. Customers can also choose their battery type (e.g. lithium ion, zinc bromide, lead acid), inverter panel size and type.
The inverter can operate in one of the three modes: automatic (explained further below), charge and discharge mode with a given rate. It is possible for commands to be sent from a central location to the inverter so that a battery charge or discharge rate can be set, or the mode of operation can be set to automatic. This can be done approximately once per minute. This is the key to the savings discussed in section \ref{sec:battery scheduling}.
A battery command schedule can be developed so that an objective function (for example, cost of electricity over the next 24 hours, or peak demand over the next month) can be minimized.
The inverter is agnostic to the battery type, but for optimal scheduling, the inverter should know the round-trip efficiency of the battery inverter combination and the state of charge of the battery at times when optimization should occur. These efficiency estimates and state of charge values are currently only known and available for lithium ion batteries from some manufacturers (e.g. Pylon, LG).
\subsection{Battery scheduling}
\label{sec:battery scheduling}
The “automatic” or default approach the inverter uses for battery scheduling is a sequence of rules applied in order. It is used for instantaneous energy management. First, load is met (in order) by solar energy, battery energy and grid energy.
After this procedure is applied, any leftover energy is sent (in order) to the battery and then the grid.
Depending on the amount of PV generated versus the amount of load in the house, the automatic approach is not necessarily the approach which minimizes the cost for the end user. The automatic approach does not consider or know about time-of-use tariffs at the household. As a selling point for the inverter, we want to lower each individual customer\textquoteright s energy bills.
In addition to the lack of knowledge about tariffs, the automatic approach does not know about the location of inverters and the associated sunrise and sunset times, weather forecast (radiation, cloud cover etc) and past load and PV data at the house.
An inverter can be commanded to maintain a charge or discharge rate or set back to ``automatic'' at any time (depending on Internet connectivity). The charge and discharge commands can be ignored if they would violate battery State of Charge (SOC) limitations. For example, the default minimum state of charge for many lithium ion batteries at customer inverters is 20\%, and a charge command is not acted upon if the battery is full.
\section{Literature Review}
Nottrott, Kleissl, and Washom (\cite{15,16,17}) and Hanna, Kleissl, Nottrott, and Ferry (\cite{10}) wrote a series of papers on different approaches used for scheduling batteries with a solar inverter. They described the OFFON, RT and OPT approaches as described below.
\begin{itemize}
\item {\bf OFFON approach:} In this approach, the battery undergoes one daily full charge cycle at 80\% depth of discharge. A constant charge rate is applied during off-peak and a constant discharge rate is applied during peak. This approach is easy to follow with a predictable load e.g. industrial or commercial, but not for a residential load.
\item {\bf RT approach:} In this approach, the battery is charged to full capacity during off-peak periods and discharged to meet customer\textquoteright s actual net load in real time. The discharge mode is the same as the automatic approach discussed in this paper.
\item {\bf OPT approach:}
This approach uses load and PV forecasts. In this model, no cost is associated with buying or selling electricity from the grid and thus time-of-use tariffs are not used. The approach is like the approach used in this paper as a linear program is used to optimize. The difference is that the function to be minimized is the sum of net PV plus battery system power output levels that fall below the forecast customer load.
\end{itemize}
In the optimized OPT approach, Nottrott et al focused on reducing monthly peak demand rather than energy cost. They performed modeling with actual load and solar forecasts at a 15-minute resolution from a commercial provider.
A so-called “trigger” was used to reassess the mode of operation if a load spike occurred in a 15-minute period or solar failed to meet prediction.
A battery command resulting in the battery level moving below the minimum State of Charge (SoC) was characterized as “failure”. For this reason, the system was deemed not yet ready for commercial operation. In contrast, in this paper, we characterize failure as depending on inverter response to commands, size of system, forecast error, PV to load ratio, and time of use tariffs. In this case, failure is when for any inverter in a set, based on a simulation compared to default mode, the automatic mode cost is lower than the “optimized” cost.
Ratnam, Weller and Kellett (\cite{18,19,20,21,22}) and Babacan (\cite{Babacan}) wrote a series of papers on scheduling residential battery storage with solar PV. They examined two approaches, 1) a linear program to minimize cost and 2) a quadratic program to reduce back-flow into distribution network in peak i.e. minimize the impact of the residential energy system on the grid. They assumed a perfect load and solar forecast, which is not realistic for an approach that needs to be implemented on real-world inverters.
The data set was from AusGrid, an Australian distribution network service provider (DNSP). It is publicly available for 300 homes and has been heavily cleaned and filtered for extreme and incorrect values.
Davy and Huang (\cite{7}) wrote a conference paper on using day ahead solar radiation forecasts to schedule a household battery.
This paper used known time of use tariffs and feed-in tariffs. Software in the MiniZinc modeling language and examples were provided. A penalty was applied for battery state of health to prevent excess battery charging and discharging to extend the battery life.
\section{problem definition}
Currently, Redback has approximately 1300 active sites across Australia and New Zealand.
Load and solar data is sent to cloud storage over wi-fi whenever a connection is available.
There are different measurements of house load and PV on the inverter for accuracy and calibration. That is, both instantaneous and cumulative values are recorded. Instantaneous values at the minute resolution reflect the average of the prior 50 seconds of data. The cumulative values are reset at midnight local time at each inverter.
The instantaneous values are measured at one-minute intervals and the internal counter for each inverter is updated at five second intervals. Data is made available and commands can be sent and responded to at a resolution of approximately one minute.
In   addition, a group of inverters can be operated as a Virtual Power Plant (VPP) or as part of an embedded network. In the case of the VPP, a group can provide demand response in the Australian National Electricity Market (NEM).
The objective for Redback is to reduce the residential energy bills for its customers. Let $I_h$ be the amount of energy imported from the power grid at a particular hour $h$ and $E_h$ be the amount of energy exported to the power grid at hour $h$. The energy bill of a particular customer is calculated as the cost of energy imported from the grid minus the revenue from energy exported to the grid. If $t_h$ is the tariff rate for energy import and $f_h$ is the rate of energy export then the energy cost at hour $h$ can be defined as:
\begin{equation}\label{eq:cost}
C_h= t_hI_h - f_hE_h
\end{equation}
Given the energy cost of each hour, the objective is to reduce the total energy cost over a given period of time. Let the given time period be represented as set of hours $T=[1 \dots H]$. The problem of reducing energy cost over {\it T} can be defined as minimizing the following function:
\[ \sum_{n=1}^H C_n\]
Ideally, we would like to have a command to charge or discharge the battery at a particular hour to minimize the overall energy cost. In the following sections, we provide details of our proposed battery scheduling approach to achieve the objective of minimizing the energy cost over a given period of time. A schematic is shown in Figure 2.
\section{Our Proposed Approach}
As shown in the equation (\ref{eq:cost}), the energy cost for each hour depends upon the tariff rates and the amount of energy transferred between the battery and the grid. While tariff rates are constant values, the values for $I_h$ and $E_h$ are unknown at the beginning of each hour. The surplus amount of energy that can be exported to the grid depends upon both the load and the PV energy generated. Hence, in order to estimate the optimal values for $I_h$ and $E_h$ for minimizing cost, we need to forecast the values for Load and PV. Next, we explain the forecasting model adopted for that purpose.
\begin{figure}[t]
	\centering
	\includegraphics[width=6.0cm]{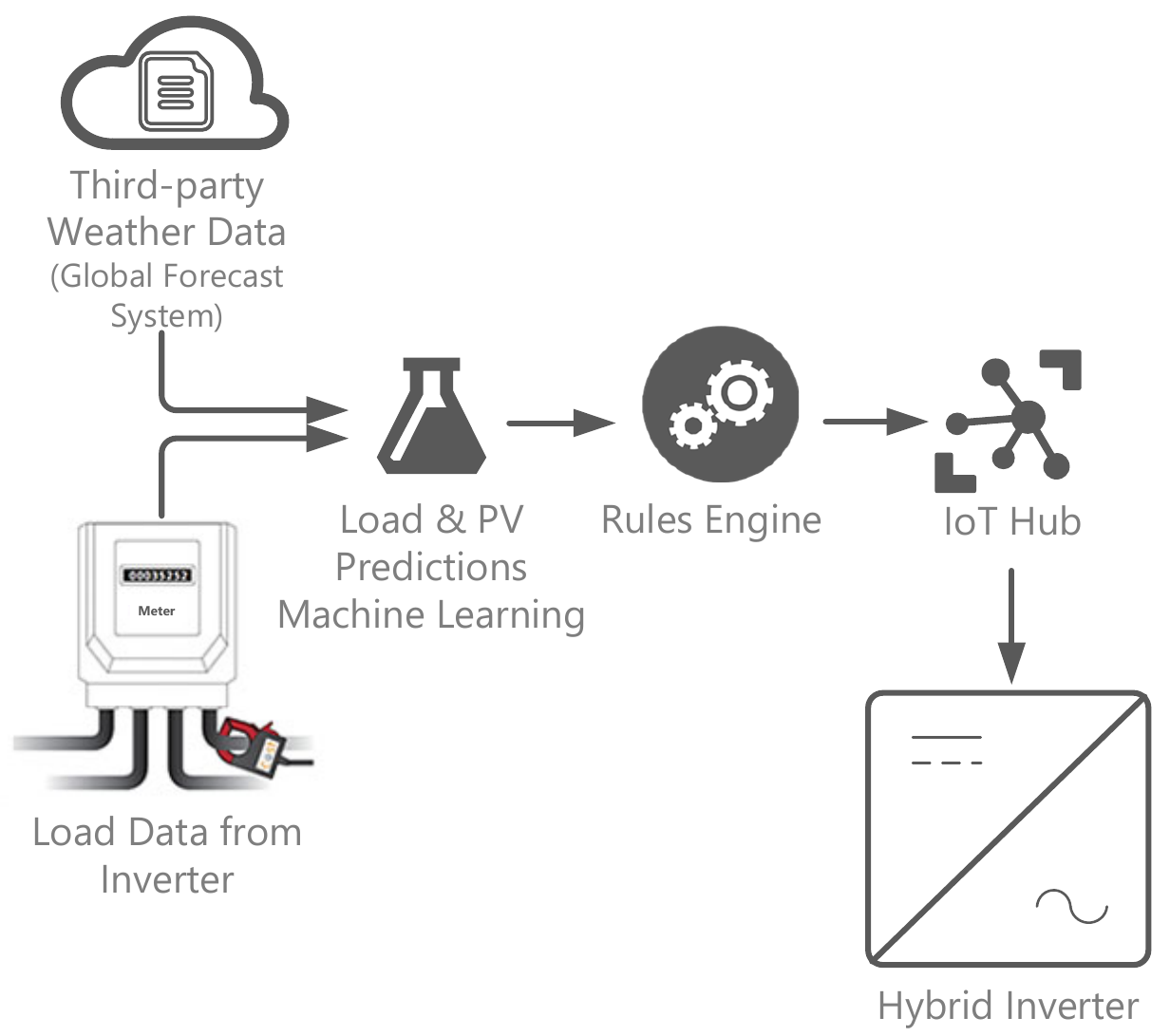}
	\caption{Schematic of prediction and optimization flow.}
	\label{fig:forecasting}	
	\vspace{-11pt}
\end{figure}
\subsection{Load and PV forecasting Model}
The amount of solar energy generated depends upon many factors like inverter size, orientation, shading effects and weather conditions. However, at Redback, the properties of each inverter (e.g. size, orientation, shading effects) are in general not known and if needed, must be derived from observations. Cloud detection and prediction e.g. using cameras at each site is not viable. Thus, we examined Numerical Weather Prediction (NWP) approaches.
Providers of NWP forecasts for Australia include the Australian Bureau of Meteorology (BOM) ACCESS model (\cite{bom}), the US National Weather Service (GFS) model (\cite{nco}), and the European Center for Medium-range Weather Forecasts (ECMWF \cite{europeancenter}).
Although access to the GFS data is free of charge, there is no “zero hour” forecast for the most relevant solar variable – downward short-wave radiation flux (DSWRF). This variable, measured in watts per square meter, provides a forecast for the solar energy for the relevant land area where the inverter is located.
To put this another way, there is no assimilation or analysis step for this variable as for the other variables. Thus the forecasts have to be used as actual data in the training step.
It is necessary to consider the spatio-temporal resolution of the NWP providers to make an informed decision.
Three organizations provide a gridded global forecast updated at six hourly intervals.
The Bureau of Meteorology (BOM) provides a global forecast, but at a finer scale, the ACCESS-C model has a temporal resolution of 1 to 36 hours (hourly) at 0.015-degree resolution for capital city areas in Australia.
ECMWF has a temporal resolution of 1 to 240 hours (3 hourly) at 0.1-degree resolution worldwide.
GFS has a temporal resolution of 1 to 384 hours (hourly to 120) at 0.25-degree resolution worldwide.
GFS also has the advantages of an online Perl script for selecting geographic subregions and variable selection, free (public domain) access and an online archive available extending back several years.
Other providers include Weather Underground, Weather Company (IBM), Weatherzone, and the Australian company Solcast (\cite{theweathercompany,wunderground,weatherzone,solcast}).
\begin{table*}[t]
	\centering
	\begin{tabular}{|p{1cm}|p{1cm}|p{1cm}|p{1cm}|p{1cm}|}
		\hline
		\multicolumn{5}{|c|}{\textbf{Percentile error values for Load and PV}} \\
		\hline
		\textbf{Percentile} & \textbf{Load NMAE}& \textbf{Load NRMSE}& \textbf{PV NMAE}& \textbf{PV NRMSE}\\
		\hline
		0\%&20\%&27\%&14\%&19\%\\\hline
		20\%&30\%&44\%&22\%&27\%\\\hline
		25\%&31\%&46\%&23\%&27\%\\\hline
		50\%&40\%&54\%&26\%&31\%\\\hline
		75\%&47\%&62\%&32\%&39\%\\\hline
		80\%&49\%&63\%&35\%&40\%\\\hline
		100\%&71\%&77\%&71\%&76\%\\\hline
	\end{tabular}
	\caption{}
	\label{tab:percentile}
\end{table*}
For each inverter location, we build a model for the hourly forecast of load and solar.
The model input data uses GFS data and solar forecasts where available. Updates are available four times daily, at 00, 06, 12, 18 UTC. As actual weather variable readings are not available for every hour, we train using the forecast values using the hourly forecasts.
The solar forecasting approach is based on the variables from the solar track in the Global Energy Forecasting Competition 2014 (\cite{GEFCom2014}).
The variables we use are: (1) temperature, (2) humidity, (3) time of day and (4) Julian date for PV. The time of day ({\it h}) and Julian date ({\it j}) are processed using sine and cosine functions i.e. $f(2\pi h/24)$ and $f(2\pi j/365)$ where $f$ is the sine or cosine function.
For solar prediction, we add variables for solar radiation, cloud cover, wind speed vectors, and atmospheric pressure. Inverse distance weighting (IDW) is used to account for points not being in the center of grid squares, as in Almeida et al (\cite{Almeida}). To summarize, the load variables are:
\begin{itemize}
	\item Sine and cosine of hour of day
	\item Sine and cosine of Julian date of year
	\item Temperature at 2 m (and IDW version)
	\item Relative humidity at 1000 hPa (and IDW version)
	\item Weekend/weekday Boolean variable
\end{itemize}
The PV variables are the Load variables plus:
\begin{itemize}
	\item Downward Short-Wave Radiation Flux (DSWRF)
	\item DSWRF leading and lagging values (at 1, 2, and 3 hours ahead and behind)
	\item Surface pressure, and differentials
	\item U and V components of wind velocity at 10 m
	\item Total Cloud Cover
	\item Wind Chill Index (Juban et al (\cite{Juban}))
	\item Solar Module Temperature (Juban et al (\cite{Juban}))
	\item Total Cloud Cover times DSWRF (Juban et al (\cite{Juban}))
\end{itemize}
We also include the IDW versions of all the variables listed above.
The model is trained hourly for load and PV using 30 days of data. We use the quantile random forest approach (\cite{R}) to generate quantile forecasts at the $40^{th}$, $50^{th}$ and $60^{th}$ percentiles. This was chosen based on the best methods in the GEFcom2014 competition adopting the random forest approach. The percentiles are used to bias the optimization which was found to save money. A more complex approach is used in Michiorri et al (\cite{Michiorri}) to account for uncertainty in the forecast.
\subsection{Accuracy of Load and PV forecasts}
\begin{figure}[t]
	\centering
	\includegraphics[trim={2.2cm 2.2cm 2.2cm 2.2cm},clip,width=8.0cm]{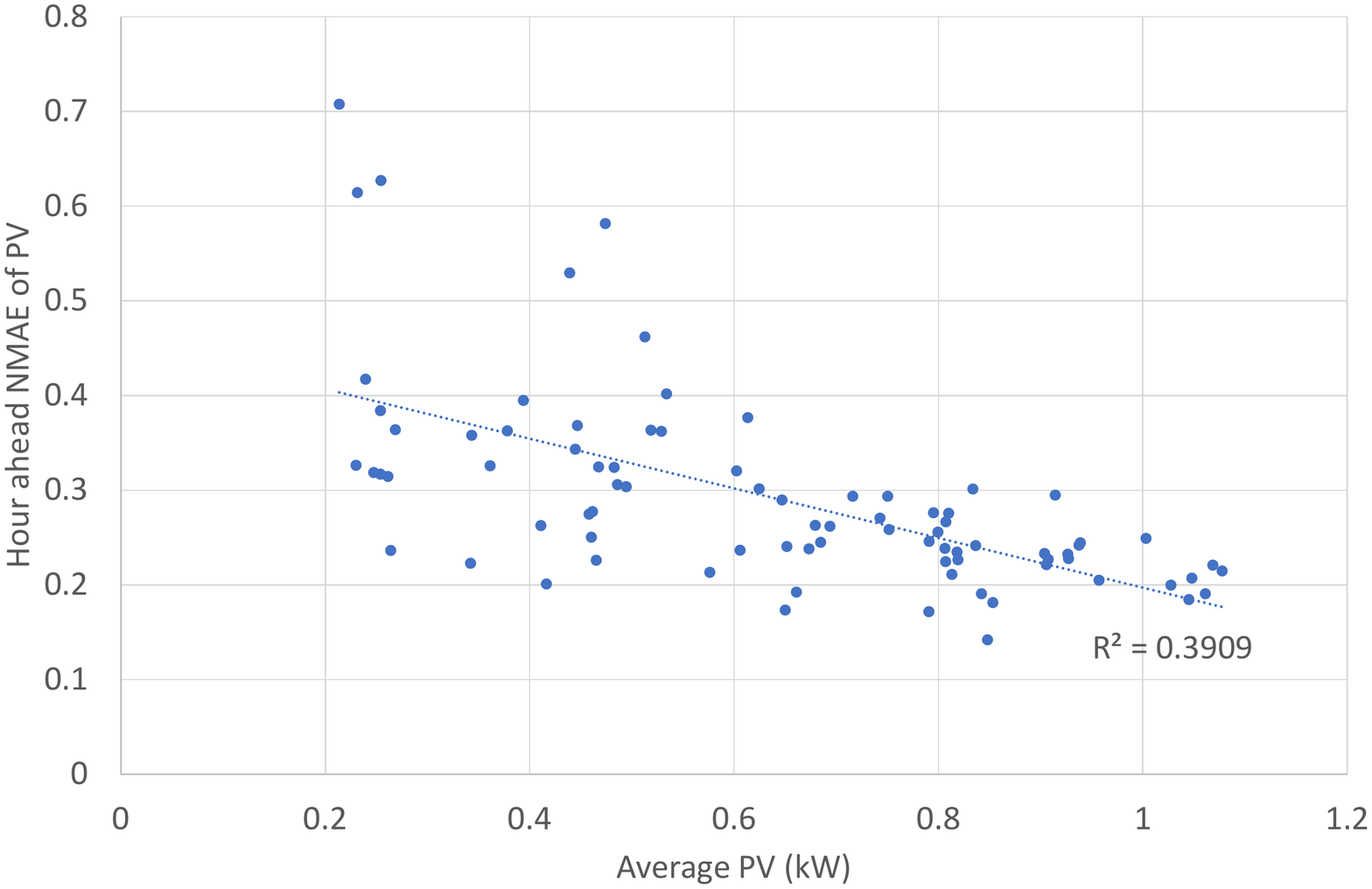}
	\caption{Mean PV vs hour-ahead NMAE by inverter.}
	\label{fig:avgPV}	
	\vspace{-11pt}
\end{figure}

The objective for Redback is to save customers money over the automatic approach while never losing money at any inverter (over the long term). Here, ``long term'' means a period of at least one year. We also write ``losing money'' referring to an outcome where the cost exceeds the automatic approach.
Forecast errors may lose money at an inverter in the short term. For example, we may run into two types of errors:
\begin{enumerate}
	\item Undercharging the battery: In case, we underestimate the load for a period, or overestimate the solar energy. Then we will not charge the battery enough during the off--peak hours. This will result in extra import from the grid during the peak hours resulting in a poor performance as compared to the automatic mode.
	\item Overcharging the battery: In case, we overestimate the load or underestimate the solar energy. We will overcharge the battery in the off-peak hours. This may result in export to the grid from full battery at the feed in tariff price.
\end{enumerate}
Clearly, the first error is more costly compared to the second error. These types of errors are much costlier at peak times when the time-of-use tariff is highest. For this reason, we run the inverter in automatic mode during peak tariff periods. In other words, instead of having the inverter follow a battery charge and discharge pattern which is believed to be optimal, the battery is scheduled in real time.
The asymmetry of the effect of forecasting errors has been noted by Khabibrakhmanov et al (\cite{Khabibrakhmanov}). The commonly-used error metric MAPE (Mean Absolute Percentage Error) is asymmetric in that it penalizes over-forecasts more than under-forecasts (\cite{Abdulla}). We measure the Mean Absolute Percentage Error for {\it n} actual values $S$ and predicted values $\hat{S}$ as:
\[
MAPE (S,\hat{S})= \frac{1}{n}\sum_{t=1}^{n}|\frac{s_t-\hat{s}_t}{s_t}|*100\%
\]
The limitation of MAPE is that error rate spikes when the actual values are near zero. Since, with PV readings the actual values are zero whenever the sun is not shining, we adopt the normalized versions of MAPE as defined by Wijaya (\cite{Wijaya}).
\[
NMAE (S,\hat{S})= \frac{\sum_{t=1}^{n}|s_t-\hat{s}_t|}{\sum_{t=1}^{n}|s_t|}
\]
\\
\[
NRMSE (S,\hat{S})= \sqrt{\frac{\sum_{t=1}^{n}\left(s_t-\hat{s}_t\right)^2}{\sum_{t=1}^{n}\left(s_t\right)^2}}
\]
We measured these error rates using the $50^{th}$ percentile predictions. Due to space restrictions, only the latest available prediction values were stored, and thus “day-ahead” error values are not available. The model was updated as often as feasible, and so at best it is “hour-ahead”.
The data shows clear patterns. For instance, the error rate for PV forecasting improves with larger PV system size as shown in Figure \ref{fig:avgPV}. Here we can see that where the mean PV is at least 1 kW, the NMAE is less than 25\%; where the mean PV is at least 600 W, the NMAE is less than 40\%, and so on.
We derived the error values in Table \ref{tab:percentile} using data from 83 inverters with between 2,208 and 5,952 hours of data available, each with at least 90\% of the hourly data present.
\subsection{Linear Programming Model}
A linear programming approach is used to determine the command to apply for charging or discharging the battery every hour when optimization is applied. The objective is to minimize the cost function defined in equation \ref{eq:cost} over a 24 hour time period. The linear program uses the following variables:
\begin{itemize}
	\item Load and solar predictions for given period (hourly). We also use the quantile forecasts to bias the input. Instead of the $50^{th}$ percentile (median forecast) load and solar, we use the $60^{th}$ percentile Load and $40^{th}$ percentile PV to avoid unnecessary grid import. That is, we overestimate the load slightly and underestimate the PV slightly, and this has been found to perform better than using the median for both.
	\item Time of use tariffs and feed in tariffs for each hour in the given period.
	\item Efficiency value for battery and inverters, as a constant value. It is more accurate to estimate efficiency as a function of ambient temperature, state of health and state of charge of the battery. This would then turn the linear program into a dynamic program. The round trip efficiency value is estimated as 84\%.
\end{itemize}
The constraints are:
\begin{itemize}
	\item Kirchhoff\textquoteright s law: Load -- PV = Battery Flow + Grid Flow
	\item State of charge limits of battery (minimum and maximum)
	\item Charge and discharge rate limits of battery and inverter
	\item Penalty values to preserve state of health of battery. In practice these are not applied as we are only applying optimization before the beginning of the peak or shoulder period.
\end{itemize}
As an output of the linear program, 24 values are generated corresponding to the each hour of the 24 hour period. Those values are then used to determine the battery flow variable for each hour in the given period. Essentially the linear program is deciding how full the battery should be, given the weather forecast for the next 24 hours. One issue here is that the linear program assumes that optimization is running for every hour in the forecast period, whereas it is transitioning to automatic mode at some points within the forecast.
\subsubsection{Linear program}
Find vectors $I=[I_h,\dots, I_{h+23}]$ and $E=[E_h,\dots E_{h+23}]$ over 24 hour time period starting from hour $h$, to minimize the function:
\[\sum_{n=h}^{h+23}t_nI_n - f_nE_n \]
subject to the following constraints:
\[ 0 \leq S_h \leq 6.5\]
\[0 \leq I_h \leq Import Limit\]
\[0 \leq E_h \leq Export Limit\]
\[-4.6 \leq B_h \leq 4.6\]
\[L_h - P_h = B_h + I_h - E_h\]
\[S_{h} = S_{h-1} -  Q_{h} * (1 + loss\_factor) + R_{h} * (1 - loss\_factor)\]
\[B_{h} = Q_{h} - R_{h}\]
where:\\
$I_h$ = import from grid in hour $h$\\
$E_h$ = export to grid in hour $h$\\
$S_h$ = state of charge at end of hour $h$\\
$B_h$ = battery flow for hour $h$ (positive is discharge, negative is charge)\\
$L_h$ = Load for hour $h$ \\
$P_h$ = PV for hour $h$ \\
$Q_h$ = Discharge for hour $h$\\
$R_h$ = Charge for hour $h$ \\
Note that all values are in kilowatt--hours (kWh) and loss\_factor = 8\% for 84\% round-trip efficiency.
The battery flow value for the first hour in the 24 hour period is used to set the battery either in the charging or discharging mode for the current hour.
A reliable connection to the central storage facility is a requirement for this solution to operate. We need a way to escape from the schedule if the Internet connection drops. Here we can revert to ``automatic'' mode if a schedule has not been received for the next hour.
\begin{table*}[t]
	\centering
	\begin{tabular}{|p{1cm}|p{8cm}|p{0.8cm}|p{0.8cm}|p{1cm}|p{0.8cm}|}
		\hline
		\multicolumn{6}{|c|}{\textbf{Tariff Descriptions}} \\
		\hline
		\textbf{Tariff}&\textbf{Description}&\textbf{Feed in Tariff (c/kWh)}&\textbf{Off-Peak Price (c/kWh)}&\textbf{Shoulder Price (c/kWh)}&\textbf{Peak Price (c/kWh)}
		\\\hline
		1&Usage - Peak usage per day - Between 7am to 11pm, Monday to Friday AEST, Usage - Off-Peak usage per day (if applicable) - All other times&11.3&23.4&&43.6\\\hline
		2&Usage - Peak usage per day - Between 7am to 11pm, Monday to Friday AEST, Usage - Off-Peak usage per day (if applicable) - All other times&11.3&20.3&&36.5\\\hline
		3&Usage - Peak usage per day - Between 7am to 11pm, Monday to Friday AEST, Usage - Off-Peak usage per day (if applicable) - All other times&11.3&20.6&&40.3\\\hline
		4&Usage - Peak usage per day - Between 7am to 11pm, Monday to Friday AEST, Usage - Off-Peak usage per day (if applicable) - All other times&11.3&21.6&&40.6\\\hline
		5&Usage - Peak usage per day - Between 7am to 11pm, Monday to Friday AEST, Usage - Off-Peak usage per day (if applicable) - All other times&11.3&18.8&&40.4\\\hline
		6&Peak Between 2pm and 8pm, Monday to Friday (excluding public holidays), Off-peak Between 10pm and 7am, Monday to Sunday, Shoulder Between 7am to 2pm and 8pm to 10pm Monday to Friday, 7am and 10pm Saturday/Sunday and Public Holidays&12.5&15.2&25&54.9\\\hline
		7&Peak 1pm to 8pm, Mon-Fri (excluding public holidays), Off Peak All other times, Shoulder 7am to 1pm and 8pm to 10pm, Mon-Fri&12.5&17.8&32.3&42.1\\\hline
		8&Peak 7am to 9am and 5pm to 8pm, Mon-Fri, Off-peak All other times, Shoulder 9am to 5pm and 8pm to 10pm, Mon-Fri&12.5&18.6&33.8&36.1\\\hline
		9&Max 7am to 9am to 5pm to 8pm, Mon-Sun, Economy All other times, Mid 9am to 5pm to 8pm to 10pm, Mon-Sun&12.5&14.4&19&27.5\\\hline
		10&Peak Between 4pm and 8pm, Monday to Friday (excluding public holidays), Off-peak Between 10pm and 7am, Shoulder Between 7am and 4pm to Between 8pm and 10pm Monday to Friday - Between 7am and 10pm Weekends&11&20.3&25.6&36\\\hline	\end{tabular}
	\caption{}
	\label{tab:tariff}
\end{table*}
\begin{table*}[h]
	\centering
	\begin{tabular}{|p{1cm}|p{1cm}|p{1cm}|p{1cm}|p{1cm}|p{1cm}|p{1cm}|p{1cm}|p{1cm}|p{1cm}|p{1cm}|}
		\hline
		\multicolumn{11}{|c|}{\textbf{Cost by tariff over 83 inverters (c/kwh)}} \\
		\hline
		\textbf{Tariff} & \textbf{No Solar} & \textbf{No Battery}& \textbf{Automatic}& \textbf{PV Persist}& \textbf{PV and Load Persisted}& \textbf{50-50}& \textbf{Load Persisted} & \textbf{60-40}& \textbf{Persist last hour}& \textbf{Perfect Forecast}\\
		\hline
		1&34.48&14.93&10.83&10.22&10.21&10.11&10.11&10.08&9.90&9.73\\\hline
		2&29.19&11.78&8.69&8.20&8.19&8.11&8.11&8.09&7.94&7.83\\\hline
		3&31.41&13.01&9.49&8.83&8.81&8.73&8.72&8.68&8.51&8.39\\\hline
		4&32.02&13.44&9.80&9.21&9.19&9.10&9.10&9.07&8.90&8.76\\\hline
		5&30.65&12.44&9.04&8.25&8.21&8.15&8.12&8.07&7.88&7.79\\\hline
		6&29.03&10.96&7.41&6.84&6.78&6.77&6.72&6.66&6.52&5.69\\\hline
		7&27.85&10.09&7.34&6.77&6.73&6.70&6.67&6.62&6.48&6.21\\\hline
		8&26.97&9.70&7.30&6.75&6.72&6.67&6.65&6.62&6.48&6.34\\\hline
		9&20.00&5.99&4.75&4.36&4.34&4.34&4.32&4.31&4.25&3.75\\\hline
		10&25.84&10.39&7.61&7.59&7.62&7.53&7.58&7.58&7.51&7.10\\\hline
		average&28.74&11.27&8.23&7.70&7.68&7.62&7.61&7.58&7.44&7.16	\\\hline
	\end{tabular}
	\caption{}
	\label{tab:savings}
\end{table*}
\begin{table*}[h]
	\centering
	\begin{tabular}{|p{1cm}|p{1cm}|p{1cm}|p{1cm}|p{1cm}|p{1cm}|p{1cm}|}
		\hline
		\multicolumn{7}{|c|}{\textbf{Number of inverters for which different approaches save money}} \\
		\hline
		\textbf{Tariff} & \textbf{PV Persist}& \textbf{PV and Load Persisted}& \textbf{50-50}& \textbf{Load Persisted} & \textbf{60-40}& \textbf{Persist last hour}\\
		\hline
		1&68&66&78&76&76&78\\\hline
		2&69&68&79&77&78&79\\\hline
		3&73&72&80&77&80&80\\\hline
		4&70&69&79&77&79&80\\\hline
		5&77&76&81&81&81&80\\\hline
		6&79&79&82&81&82&82\\\hline
		7&76&75&80&79&80&80\\\hline
		8&74&74&80&79&80&79\\\hline
		9&83&83&83&83&83&83\\\hline
		10&36&28&51&37&36&53\\\hline
		Total&705&690&773&747&755&774 \\\hline
	\end{tabular}
	\caption{}
	\label{tab:inverters}
\end{table*}
\section{Evaluation}
We simulated the optimization over 83 inverters (with an average of about 4,000 hours of data each) with an assumed round-trip efficiency value of 84\%, battery storage of 6.5 kWh, and battery import and export limits of 4.6 kWh/h. The tariff rates used are presented in Table \ref{tab:tariff}. It is important to carefully select the inverters with good quality data to participate in the simulations. Therefore, we selected the inverters using the following criteria:
\begin{enumerate}
	\item {\bf PV and Load mean values}: As the resolution of the hourly load data is only 0.1 kWh, or an average of 100 W per hour, we restrict the algorithm to consider only those with an average PV and average load of at least 200 W. For inverter data up to 18 July 2018 with at least 2,160 hours of readings, 611 of 679 inverters have an average load of at least 200 W, 606 of 679 inverters have an average PV of at least 200 W, and 579 of 679 (85\%) meet both criteria. The best performing 1.5 kW solar panels in Australia achieve an average PV output of 300 W, but the Redback inverter can invert up to 4.6 kWh of solar per hour so is more suited to large solar panels.
	\item	{\bf PV to Load ratio:} Of the 83 inverter data set, the median PV/Load ratio is 0.89 with $25^{th}$ percentile 0.66 and $75^{th}$ percentile 1.29.
	\item {\bf Time zone boundary of tariffs versus forecast period}: With GFS data, extra coding is required to account for time zones which are on the half hour boundary. For example, South Australia and the Northern Territory is GMT+9:30 or GMT+10:30 depending on daylight saving time, whereas the eastern states of Victoria, New South Wales, Queensland, and Tasmania, plus Western Australia, are aligned with hourly boundaries relative to GMT (GMT+8, 9, 10 and 11). Currently, time-of-use tariffs are not yet available in South Australia or the Northern Territory.
	\item {\bf Forecast error rate (e.g. MAPE)}: If for some reason the forecast error rate at a site is high, this may indicate problems with the data at that location, possibly due to poor network connectivity, or that the period of available data is too short to obtain reasonable forecasts.
	\item {\bf “Trigger” approach}: Hanna et al (\cite{10}) present the use of a “trigger” and Bergner and Quaschning (\cite{Bergner}) proposed a “real time correction” algorithm. Those techniques adjust the charge and discharge values during an optimization period in order to meet a target. In our case, we may choose to revert to automatic mode if a target is missed.
	\item {\bf Actual time of use tariffs in effect}: There is no advantage in load shifting or charging the battery from the grid at certain times of the day if the tariff is flat throughout the day.
	\item {\bf Credibility of data}. Some customers turn off their wi-fi overnight and thus some hourly load data is unavailable. There are also data quality issues with some of the PV and Load data which must be assessed.
\end{enumerate}
Table \ref{tab:savings} shows the cost over all the selected inverters in units cost per kilowatt hour (c/kWh). We assess the cost for the following ten cases:
\begin{enumerate}
	\item No solar system available
	\item No battery available (by netting solar from the load, with no losses assumed)
	\item Automatic mode
	\item Predictions with each hourly PV value taken to be the same as 24 hours previously (persistence). The first 24 hours are taken to be a perfect forecast.
	\item Persistence with each hourly PV and load the same as 24 hours previously
	\item Predictions with $50^{th}$ percentile PV and load
	\item Predictions with each hourly Load the same as 24 hours previously
	\item Prediction with $60^{th}$ percentile load and $40^{th}$ percentile PV
	\item Persistence with hourly PV and load the same as one hour previously
	\item Perfect forecast
	
\end{enumerate}
\subsection{Cost Savings}
The results of our simulations show that our proposed approach is successful in reducing the cost of the energy bills. Over the entire simulation period  for each inverter, using hourly simulation and over ten different tariffs, the optimization saves 1-10\% over the automatic approach, compared to 7-23\%, if a perfect PV and load forecast was available. Table \ref{tab:inverters} shows the number of inverters for which different approaches save money over the automatic approach. In practice, it seems the “50-50” approach saves money at the most inverters, at the cost of a slightly more expensive average cost compared to “60-40”.
The simulation results show that the persistence approach for predictions (of 24 hours ago) results in worse outcomes for the optimization. For every tariff, the load only persistence forecast performs worse in terms of the number of inverters with a loss and the mean saving, and the same pattern applies moving from the load only persistence forecast to the load and PV persistence forecast. This is in line with the results of Weniger et al (\cite{Bergner}).
\begin{figure}[h]
	\centering
	\includegraphics[trim={2.2cm 2.2cm 2.2cm 2.2cm},clip,width=8.0cm]{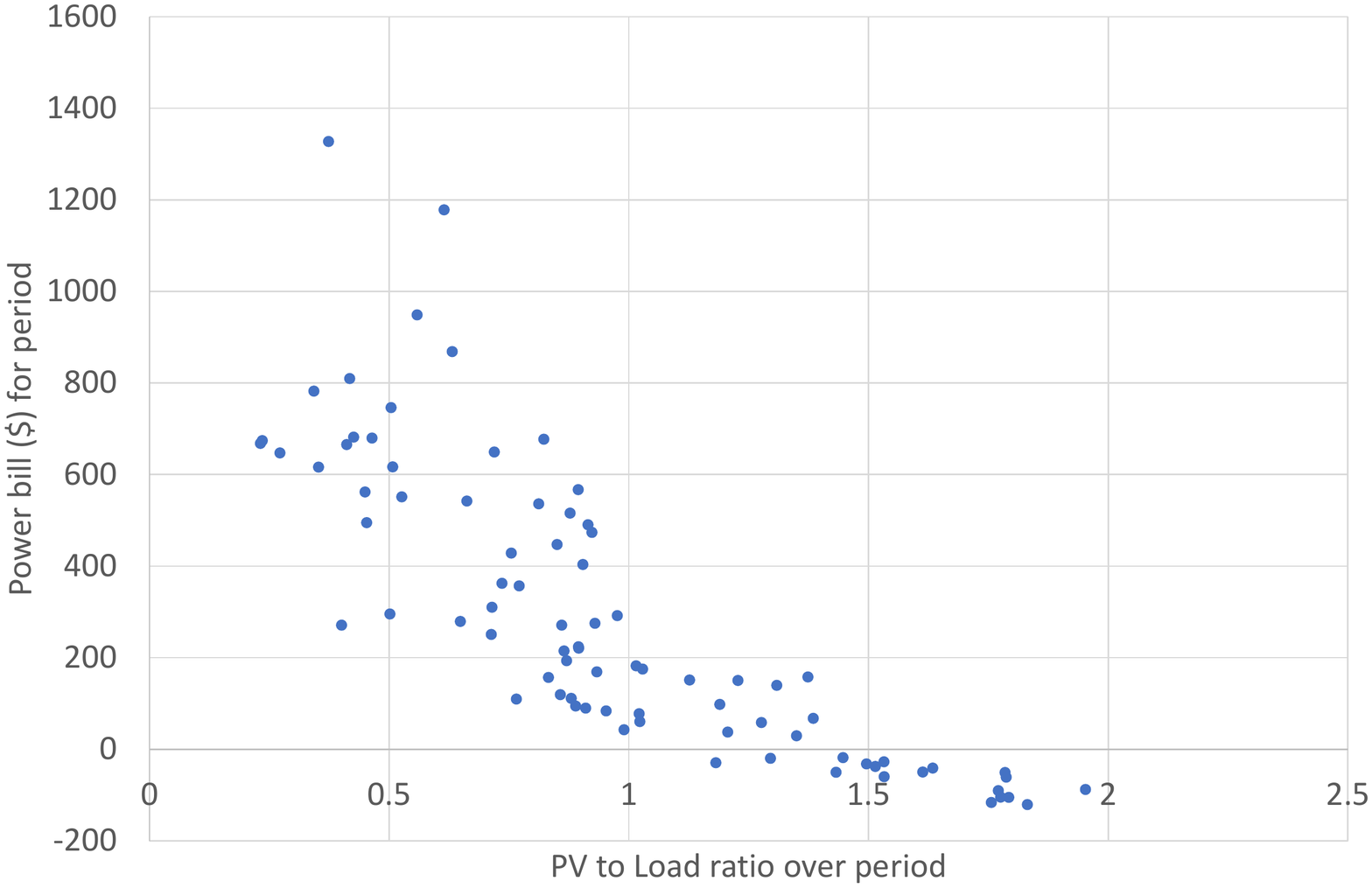}
	\caption{Bill with perfect forecast vs PV/Load ratio of data (Tariff 1).}
	\label{fig:pvtoload}	
	\vspace{-11pt}
\end{figure}
\subsection{Impact of Tariff Price}
\begin{figure}
	\centering
	\includegraphics[trim={2.2cm 2.2cm 2.2cm 2.2cm},clip,width=8.0cm]{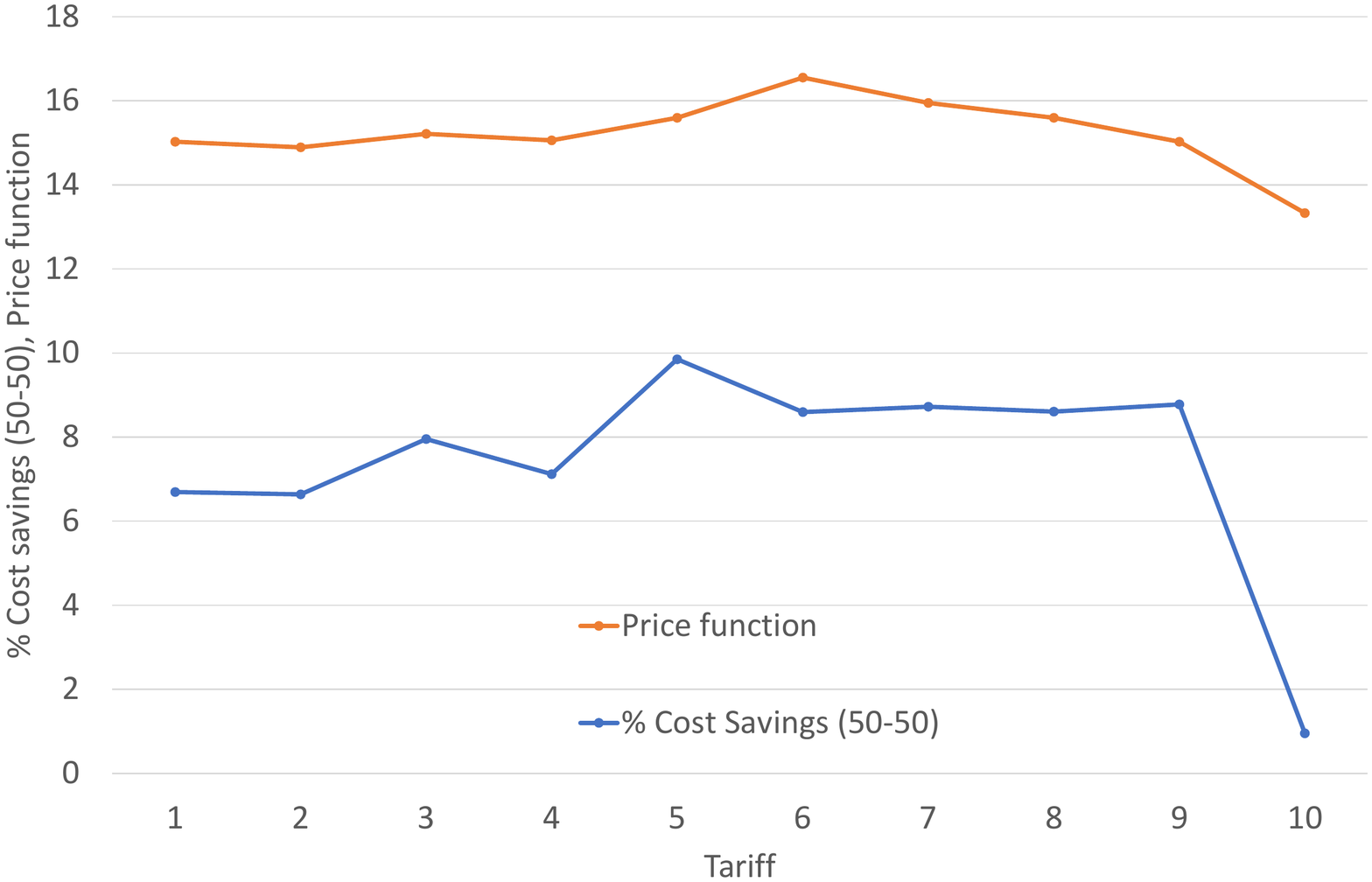}
	\caption{Impact of feed--in tariff and peak plus shoulder to off--peak price ratio}
	\label{fig:tariff}	
	\vspace{-11pt}
\end{figure}

In simulations it is observed that, generally, with a time--of--use tariff where the peak price is very high relative to the off--peak, it is optimal to attempt to fully charge the battery before the peak begins, with a combination of grid and solar depending on how much solar is forecast. Here, we only apply optimization before peak on weekdays.
It is also observed that the percentage cost savings over automatic mode are a function of both the feed--in tariff price (the percentage is higher for higher feed--in tariff prices) and the ratio of the sum of the peak plus shoulder hourly prices to the sum of the off-peak hourly prices (the percentage is higher for higher ratios). Figure \ref{fig:tariff} shows the strong correlation between cost savings and this function, that is, the sum of the feed--in tariff price and the ratio of peak plus shoulder to off--peak prices.
\subsection{Impact of PV to Load ratio}
On days where PV to Load ratio is very low (for example, less than 0.5) the automatic approach will not be optimal, as the optimal approach may charge the battery from the grid during off-peak times and discharge the battery during peak times. However, the automatic approach will never charge the battery from the grid.
Where the PV to Load ratio is very high (for example, more than 2.0) it is not possible to save significant money over the “theoretically perfect forecast” optimal schedule as the “automatic” approach is very similar. Where the average PV or Load value is too low, the measurement error and variability in the values becomes too high to produce an optimal schedule.
For this reason, we only apply the optimization where the average PV and Load values are at least 200 W and the inverter\textquoteright s long-term PV to Load ratio is less than 1. In general, when the long-term PV to Load ratio is above this number the costs per inverter are negative as the feed-in tariff price received for PV energy exported to the grid exceeds the charge for any energy imported from the grid. As seen in Figure \ref{fig:pvtoload}, above the given ratio it can be difficult even to find a consistent strategy to save money over automatic mode.
\section{Conclusions}
Further research will include integrating other forecasting approaches. It is known that ensembles of weather forecasting services do better than individual services (Ren et al (\cite{Ren})).
We can improve forecast accuracy of solar and load by using solar data providers such as Solcast (\cite{solcast}) which use multiple information sources (GFS, ACCESS, ECMWF, and Himawari-8 satellite data).
We can use the simulation results developed to recommend battery and panel size to new customers. For example, the batteries may be available in increments of 3.3 kWh up to 13.2 kWh, and solar panels may be available in different sizes.
We can develop a `trigger' or `real time correction' mechanism, although this is less relevant in our case as the optimization is done early in the morning and the time resolution of load and PV forecasting should improve over time. In our simulation we assume the charge and discharge commands can always be obeyed as the grid import or export value is determined by the actual PV and load values.
The state of health of the battery should be considered in the linear program. Other research (for example, Davy and Huang in \cite{7}) contains a penalty against cycling the battery unnecessarily to avoid aging the battery prematurely.
Battery efficiency can be estimated more effectively. Currently this is assessed as a constant value and this has the advantage that the optimization is a simple linear program which can be run quickly.
Avoiding back flow to grid may be a concern in future. This is a key theme of the Ratnam et al papers as these authors are considering the issues from the perspective of a Distribution Network Service Provider (DNSP) in Australia, Ausgrid.

\end{document}